\documentclass[aps,prb,preprint,showpacs,groupedaddress]{revtex4}
\usepackage[dvips]{graphicx}
\begin{document}
\title{Unusual transport properties of ferromagnetic Heusler alloy Co$_2$TiSn}
\author{S. Majumdar}
\author{M. K. Chattopadhyay}
\author{V. K. Sharma}
\author{K. J. S. Sokhey}
\author{S. B. Roy} 
\author{P. Chaddah}
\affiliation{Low Temperature Physics Laboratory, Centre for Advanced Technology, Indore 452013, India.}
\date{\today}
\begin{abstract}
We report results of magnetization, zero field resistivity and magnetoresistance measurements in ferromagnetic Heusler alloy Co$_2$TiSn. There is a striking  change in the character of electron transport as the system undergoes the paramagnetic to ferromagnetic transition. In the paramagnetic state the nature of the electron transport is like that of a semiconductor and this changes abruptly to metallic behaviour at the onset of ferromagnetic ordering. Application of external magnetic field tends to suppress this semiconducting like transport leading to a negative magnetoresistance which reaches a peak in the vicinity of Curie temperature. Comparison is made with the similar unusual behaviour observed in other systems including UNiSn and manganites.
\end{abstract} 
\pacs{75.30.Kz}
\maketitle

In recent years recognition of half heusler compound NiMnSb as a potential spin-injector material for spintronics applications\cite{1} and the discovery of large magnetic shape memory effect in the full Heusler compound Ni$_2$MnGa\cite{2} have stimulated much research activity on Heusler alloys in general. The full Heusler alloys have composition X$_2$YZ forming in L2$_{1}$ structure while the half Heusler alloys with composition XYZ forms in C1$_b$ structure. Here X and Y are transition elements and Z is an sp-element. Traditionally compounds with Mn occupying the Y-site have been drawing the most of the attentions as they form ideal systems for studying localized 3d metallic magnetism, and such studies of course revealed various interesting functionalities in these materials. Non manganese Heusler alloys are interesting too. For example Fe$_2$VAl\cite{3} and Fe$_2$TiSn \cite{4} have been subject of much attention since 1990s due to the possibility of heavy fermion behaviour in them. On the other hand Co-based Heusler alloys Co$_2$YZ, where Y=Ti,Zr, Hf etc and Z=Sn or Al, are considered to be good candidates for studying itinerant electron ferromagnetism\cite{5}. Of these Co$_2$TiSn is particularly interesting because of its similarity with the prototype half metallic system NiMnSb\cite{6}. The shape of the spin-polarised total density of states and the dispersion curves for Co$_2$TiSn resembles the electronic spectra of half-metallic ferromagnets \cite{7}. Also the magnetic moment per formula unit is close to an integer number (1 $\mu_B$), which should be the case for half-metallic systems. Here we present results of magnetization, resistivity and magnetoresistance study in a polycrystalline sample of Co$_2$TiSn. The resistivity shows metallic behaviour below the paramagnetic(PM)- ferromagnetic(FM) transition temperature (T$_{Curie}\approx$ 355K). This is consistent with earlier resistivity studies on Co$_2$TiSn between 4 and 300K \cite{8}. The nature of the resistivity behaviour, however, changes from metallic to semiconductor-like at T$_{Curie}$. To the best of our knowledge this metal-semiconductor like transition in Co$_2$TiSn has not been reported so far. We show that the resistivity around T$_{Curie}$ is quite sensitive to applied magnetic field. Anomalous negative coefficient of resistivity above the magnetic ordering temperature has earlier been observed in various (Fe$_{1-x}$M$_x$)$_3$Ga (M=V and Ti) and (Fe$_{1-x}$N$_x$)$_3$Si (N=V, Mn, Ti and Cr) based pseudobinary alloys\cite{9,10} and UNiSn \cite{11}. We shall compare our results with these earlier experimental findings. Further we discuss our results in the light of a fairly recent theory on the resistivity due to spin-dependent scattering of carriers in  ferromagnetic metals with localized spins\cite{12}.   

Polycrystalline samples of Co$_2$TiSn are prepared by argon-arc melting from starting  materials of 4N purity. The sample has been annealed in vacuo at 800$^0$C for 7 days. The sample is characterized with standard X-ray diffraction study and it consists of a single phase with cubic L2$_1$ Heusler structure. Magnetization measurements are performed with a commercial SQUID magnetometer (Quantum Design MPMS-5). Resistivity of the sample is measured with the standard four probe technique. A home made cryocooler is used for zero field resistivity measurements in the temperature regime 30K$\leq$T$\leq$385K. A separate oven system is employed to extend the temperature range to 450K. A superconducting magnet with field up to 80 kOe is used for magnetoresistance study. 

\begin{figure}[t]
\centering
\includegraphics[width = 7 cm]{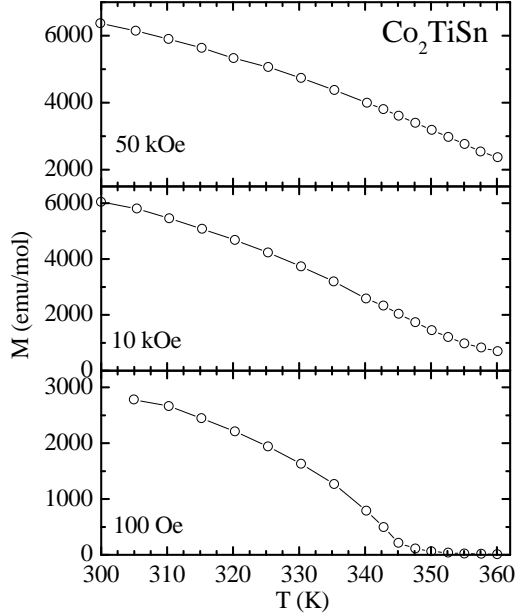}
\caption{Magnetization versus temperature data of Co$_2$TiSn for different applied fields. Data exist down to 5K but not shown here for the sake of clarity. }
\label{mt}
\end{figure} 

Fig.1 presents the magnetization (M) versus temperature (T) plot for the polycrystalline Co$_2$TiSn sample in  fields of 100 Oe, 10 kOe and 50 kOe. The T$_{Curie}$ is estimated from the inflection point in the M vs T curve (or even better from the peak of dM/dT vs T curve). The estimated T$_{Curie}$ from the 100 Oe M vs T data is  $\approx$350K. This is slightly lower than the reported value of 359K\cite{13}.However, it is clear from Fig.1 that with the increase in the applied field the ferromagnetism continues to persist in the higher T regime. Consequently the estimated values of T$_{Curie}$ are higher. These results clearly show that the magnetic properties of Co$_2$TiSn around the PM-FM transition are quite sensitive to the applied field.

Fig.2 presents the resistivity ($\rho$) versus T plot in zero applied field. There is a distinct change in the resistivity from metallic to semiconductor-like taking place at 355K. This temperature matches well with the T$_{Curie}$ determined from the magnetization measurements, hence we comment that this change in the character of electronic transport is associated with the FM-PM transition. Although the resistivity in the FM regime shows metallic character, magnitude of the resistivity remains fairly high (above 300 $\mu \Omega$-cm) even in the low temperature regime. However, it was possible to fit this metallic resistivity with a standard Bloch-Gruniessen law plus a T$^2$ term. The Debye temperature estimated from this fitting is 411K which is comparable to that of the isostructural Heulser compound Fe$_2$TiSn \cite{4}. The T$^2$ term can arise due to the spin-flip scattering of charge carriers by magnons in a ferromagnet, and also due to strong interactions in the Fermi liquid. A careful study of resistivity, at least down to helium temperature,  is needed to discern amongst the various possible origins. Such a study is in progress now.  On the other hand, the resisvity above 355K can be fitted into an expression e$^{Eg/2k_BT}$ representing activated behaviour in a semiconductor. Fig.3 presents logarithm of resistivity versus 1/T plot and a band gap of 12.3$\pm$1 mev is estimated from the slope of this plot. 

\begin{figure}[t]
\centering
\includegraphics[width = 7 cm]{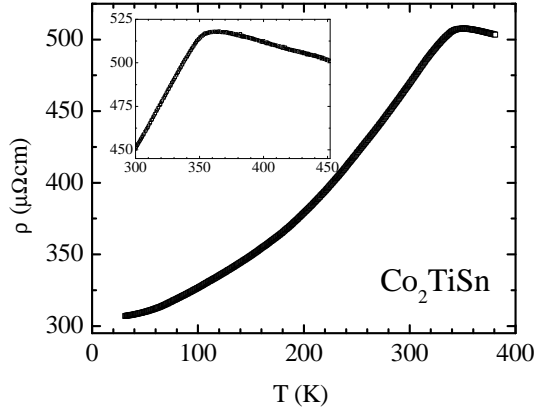}
\caption{Resistivity behavior of Co$_2$TiSn as a function of temperature. The inset shows an enlarged view of the high temperature data (300-450 K) measured in a oven.}
\label{rt}
\end{figure} 

Fig. 4 presents the effect of applied magnetic field on the resistivity in and around the PM-FM transition region. The zero field semiconducting like behaviour tends to get suppressed with the increase of the magnetic field. For the sake of clarity we present in Fig.4 only the results with extreme fields of 0 and 80 kOe. Our preliminary study indicates that the semiconducting gap is reduced by 10\% with the application of 10 kOe field. However, we need to extend the present temperature range of field dependent resistivity measurement to make a firm comment on the field dependence of the semiconducting gap.  In Fig.5 we plot  magnetoresistance obtained from the results presented in Fig.4. Magnetoresistance shows a distinct extremum around T$_{Curie}$ which suggests a correlation between the spin orientation and the electron scattering. This conjecture is further supported by the observed change in the temperature dependence of magnetization with H in the same T regime (see Fig.1). We have also measured magnetoresistance as a function of field at various fixed temperatures between 50 and 380K. Magnetoresistance becomes negligibly small below 200K and above 370K. 

\begin{figure}
\centering
\includegraphics[width = 7 cm]{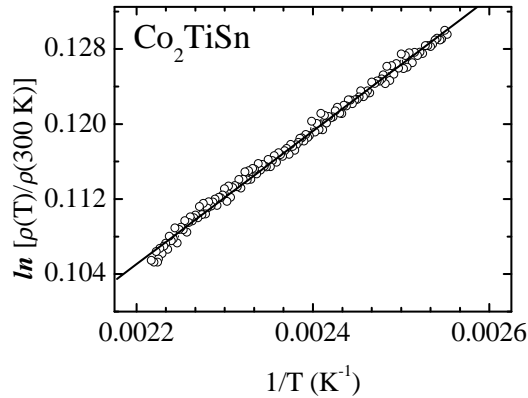}
\caption{Natural logarithm of the normalized resistivity as a function of inverse temperature. The solid line is a linear fit to the data. The temperature range covered is 380K to 450K.}
\label{lnr}
\end{figure}

A negative coefficient of resistance has earlier been reported above the Curie temperatures for Ni-doped Co$_2$TiSn alloys\cite{8}. In that work the resistivity measurement for pure Co$_2$TiSn, however, was up to 300K only, hence only metallic behaviour was reported for this parent compound. It is well known from the band structure calculation that the spin down density of states function of Co$_2$TiSn varies strongly with energy at the Fermi level (E$_F$). As a consequence a slight change in the position of the Fermi level can cause significant variations in the transport properties and this aspect was investigated in details with doping studies\cite{8}. However, no definite explanation was provided for the observed negative coefficient of resistance in the Ni-doped Co$_2$TiSn.

Such unusual correlation between semiconductor-metal transition and magnetic transition has earlier been observed in UNiSn and this has been a subject of considerable experimental\cite{11} and theoretical\cite{14,15} attention. UNiSn, like its isostructural compound NiMnSb, is considered to be a half-metallic ferromagnet\cite{14,15,16}. From the resistivity measurements an energy gap of 120 mev has been estimated for the semiconducting state of UNiSn. It is conjectured that at the onset of the magnetic transition  the energy gap of at least one of the spin bands disappears leading to the metallic behaviour in resistivity. This picture is consistent with the half metallic character of the magnetic state. These results can prompt to think of a one-to-one correspondence between UNiSn and Co$_2$TiSn. However, some notes of caution here: first the half metallic nature of Co$_2$TiSn is yet to be established firmly, and second there remains some controversy on the magnetic ground state of UNiSn whether it ferromagnetic\cite{11} or antiferromagnetic\cite{15,16}.

\begin{figure}[t]
\centering
\includegraphics[width = 7 cm]{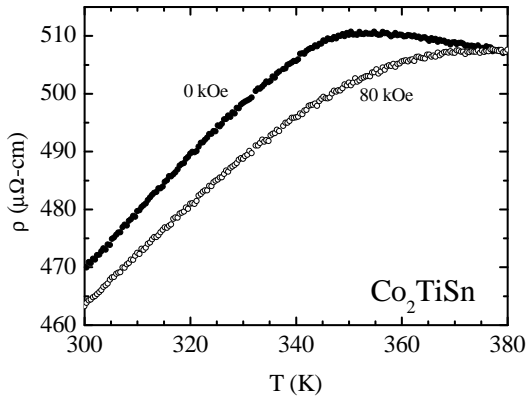}
\caption{Resistivity versus temperature plot measured at zero and 80 kOe magnetic fields.}
\label{rth}
\end{figure} 
     
It is interesting  to note here that the electronic structure calculations in the sister compound Fe$_2$TiSn predicted a nonmagnetic ground state and the existence of pseudogap at the Fermi level\cite{4}. In real systems, lattice disorder complicated the possibility of a metal-semiconductor transition; a weak ferromagnetic ground state was observed below 240 K followed by semimetal to semiconductor like transition around 50K \cite{4}. In addition there is evidence of heavy fermion behaviour in Fe$_2$TiSn in the low T regime \cite{4}. Effect of Co-doping in Fe$_2$TiSn has been studied in details and a negative coefficient of resistivity has been reported for FeCoTiSn for a substantial temperature regime starting from room temperature \cite{17}. However, no definite explanation is available for such behaviour.  It is also interesting to note that the Heusler alloy Fe$_2$VAl which is also on the verge of a magnetic ordering shows semiconducting behaviour  down to 2K \cite{3}. While initial specific heat measurements\cite{3} indicated the possibility of electronic mass enhancement in this compound, latter studies suggested that sample-dependent Schottky anomaly originating from magnetic clusters associated with the Fe-defects might be at the origin of the anomalous specific heat behaviour \cite{18}.

A negative coefficient of resistivity above Curie temperature has earlier been reported for  ferromagnetic systems like V and Ti-doped Fe$_3$Ga \cite{9}, V, Ti, Mn and Cr-doped Fe$_3$Si pseudobinary alloys\cite{10}. A combination of small conduction electron number per atom and very large spin-disorder scattering was proposed to be the cause of such anomalous resistivity behaviour above the Curie temperature. It was experimentally observed that only those alloys with the resistivity in the paramagnetic state above 150 $\mu\Omega$-cm showed the unusual resistivity  behaviour \cite{10}. We note here that the resistivity value of $\approx$ 500 $\mu\Omega$-cm in the paramagnetic state of Co$_2$TiSn fulfills this criterion.  A more recent experiment on the transition metal compound MnRhP has revealed all these interesting features \cite{19}. 

\begin{figure}[t]
\centering
\includegraphics[width = 7 cm]{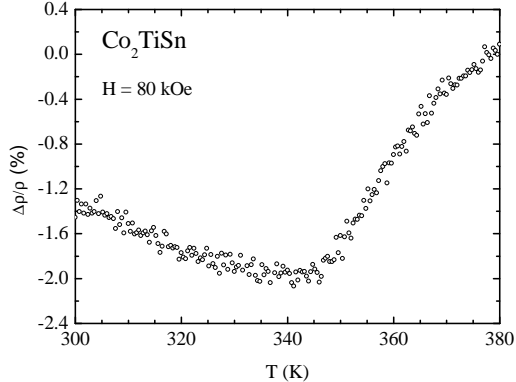}
\caption{Magnetoresistance ($\Delta \rho/\rho$) as a function of temperature at 80 kOe of applied magnetic fields. Here $\Delta \rho$ is defined as [$\rho$(80 kOe)- $\rho$(0 Oe)] and $\rho$ is $\rho$(0 Oe).}
\label{mr}
\end{figure}

Changing sign of the coefficient of resistivity at the PM-FM transition is one of the remarkable features of many manganite compounds showing colossal magneto resistance (CMR). The magnitude of the resistivity change and its sensitivity to applied H around T$_{Curie}$ is very high in CMR-manganites, hence leading to the drastic CMR effects \cite{20}. While we note that the magnetoresistance in Co$_2$TiSn is definitely not comparable with the CMR observed in various manganites, qualitatively the observed peak in magnetotransport around T$_{Curie}$ in Co$_2$TiSn is very similar in nature.  Magnetic interactions and underlying microscopic origin of the magnetic phase transition in Heusler alloys are expected to be quite different from those in manganites and underlying connections between these two different classes of materials are not very obvious. 
However, there is fairly recent theoretical effort to understand such behaviour in terms of spin-dependent scattering of carriers in ferromagnetic materials with localized spins \cite{12}. In this theoretical framework carrier concentration is a key factor and so is the spin fluctuation especially when a ferromagnetic state is almost unstable against another magnetic state. This theory suggests that in the systems with small Fermi surface, critical spin fluctuations with long wavelengths can contribute to the resistivity to give rise to a peak at T$_{Curie}$.  This peak becomes more sharp in a metal whose ferromagnetic state is on the verge of an instability.  External magnetic fields can easily suppress critical spin fluctuations with long wavelengths, resulting in a negative magnetoresistance. A ferromagnetic metal with a small Fermi surface can often become half-metallic because of splitting of the up and down-spin bands due to magnetization. While Co$_2$TiSn has a low carrier concentration and it is supposedly a half metallic ferromagnet, it is not very clear to the present authors whether detail characteristics of the Fermi surface of Co$_2$TiSn would really fit within such theoretical picture. 

In conclusion, combining the results of magnetic and transport measurements we show that Co$_2$TiSn undergoes a semiconductor-metal transition around 350K and this is associated with a PM-FM transition. In the light of the existing results on UNiSn and recent theoretical studies on spin-dependent scattering of carriers in ferromagnetic materials, the metallic behaviour in the ferromagnetic state of Co$_2$TiSn is consistent with the conjectured half-metallic character of this compound. It seems that Co$_2$TiSn belongs to a growing class of ferromagnets with low carrier concentration which shows such unusual correlation between magnetic ordering and metal-semiconductor transition.

{99}                        

\end{document}